\documentclass[conference,a4paper,flushend]{iaria} 
\usepackage{caption}
\usepackage{tabularx} 

\title{Assessing the Carbon Footprint of Virtual Meetings: A Quantitative Analysis of Camera Usage}
\author{
  \IEEEauthorblockN{%
    Félix Mortas\,}
  \IEEEauthorblockA{%
    Independent Researcher\\
    Paris, France\\
    e-mail: {\tt \href{mailto:felix.mortas@hotmail.fr}{felix.mortas@hotmail.fr}}
} }
\begin{document}
\maketitle
\begin{abstract}
This paper quantifies the carbon emissions related to data consumption during video calls, focusing on the impact of having the camera on versus off. The findings regarding the environmental benefits achieved by turning off cameras during meetings challenge the claims of some prevalent articles. The experiment was carried out using a 4G connection via a cell phone to measure the varying data transfer associated with videos. The outcomes indicate that turning the camera off can halve data consumption and associated carbon emissions, particularly on mobile networks. The paper concludes with recommendations to optimize data usage and reduce the environmental impact during calls.
\end{abstract}

\section{Introduction}
The Information and Communication Technology (ICT) sector is responsible for approximately 2 to 4\% of global greenhouse gas (GHG) emissions~\cite{Freitag2021}. With the advent of generative AI, reputable institutions project a significant increase in these emissions in the near future~\cite{Iea2025}. To mitigate the environmental impact of this growth, both IT developers and end-users are increasingly adopting responsible practices.

A prevalent assertion suggests that deactivating cameras during virtual conferences may mitigate GHG emissions.
Few serious studies have been conducted to determine the extent to which this best practice reduces environmental impact.
The most popular and serious study on this subject highlights a 96\% reduction in environmental impact by turning off cameras during virtual meetings~\cite{Obringer2021}.
However, while the findings are promising, they are largely based on estimates and modeling rather than direct experimental data.
To obtain more accurate and actionable insights, further studies that include real-world experiments are necessary.

This report quantifies and analyzes the data consumption during video calls, focusing on the impact of having the camera on versus off. The analysis includes a preliminary statistical test and recommendations based on the findings.

This paper is structured as follows. Section II synthesizes existing research on the energy intensity of fixed and mobile network infrastructures. Section III describes the methodology used for the study. Section IV presents the results. Section V discusses the interpretation of the results and offers practical recommendations for digital sobriety. Finally, Section VI summarizes the key findings and suggests avenues for future research.

\section{State of the Art}
The consumption of 4G data is closely linked to carbon dioxide equivalent (CO$_2$e) emissions. The energy consumption of mobile networks such as 4G varies significantly according to the volume of data transferred. According to some estimates, transferring one gigabyte (GB) of data on 4G corresponds to a consumption of electricity of 0.6~kWh~\cite{Arcep2019} which contributes directly to CO2 emissions associated with the use of these technologies.

While national regulators provide foundational estimates, the energy intensity of data transfer is subject to significant fluctuations based on infrastructures and methodological choices made in each study.

\begin{table}[htbp]
\caption{Energy consumption of different mobile networks.}
\label{tab:mobileenergyconsumption}
\centering
\begin{tabular}{|l|l|l|l|}
\hline
\textbf{Network} & \textbf{kWh/GB} & \textbf{Location} & \textbf{Year} \\
\hline
Mobile & 0.14 & Europe & 2020 \\
\hline
4G & 0.17 & Finland & 2020 \\
\hline
4G & 0.26 & Belgium & 2020 \\
\hline
Mobile (3G, 4G, 5G) & 0.12 & Finland & 2022 \\
\hline
\end{tabular}
\end{table}

Table~\ref{tab:mobileenergyconsumption} presents a comparative analysis of energy consumption in kilowatt-hours per gigabyte (kWh/GB) using various mobile networks across different locations and years.

\textcite{EuropeanCommission2020} reports an energy consumption of 0.14~kWh/GB in Europe for the year 2020 for mobile network, not precising the technologie used. In Finland, \textcite{Nokia2025} indicates that the 4G network consumes 0.17~kWh/GB. \textcite{Golard2020} presents data for Belgium, where the 4G network consumes 0.26~kWh/GB in 2020. Additionally, \textcite{Traficom2022} provides a comprehensive analysis for Finland, showing that mobile networks, including 3G, 4G, and 5G, consume 0.12~kWh/GB in 2022.

In contrast, fixed networks such as fiber optics and ADSL have a different energy profile. According to a statement by the ARCEP French organization, the energy consumption of fixed networks depends relatively little on how they are used. For example, fiber optics consume an average of 0.5~W, with little variation, regardless of the amount of data transmitted~\cite{Arcep2019}. The energy efficiency of fiber networks is due to the physical advantage of light that does not require active network elements, as well as the high data rates allowed by the technology. The largest consumer of electricity in fixed networks is the customer premises equipment (CPE), including a modem router unit and a possible termination of the optical network~\cite{Prysmian2021}. This means that even with intensive use, the impact on electricity consumption remains marginal.

Although fairly stable, data on electricity consumption related to data transfer over fixed networks have been published in the literature.

\captionsetup{font=footnotesize,justification=centering,labelsep=period}

\begin{table}[htbp]
\caption{Energy consumption of different fix networks.}
\label{tab:fixenergyconsumption}
\centering
\begin{tabular}{|l|l|l|c|}
\hline
\textbf{Network} & \textbf{kWh/GB} & \textbf{Location} & \textbf{Year} \\
\hline
Fix (ADSL) & 0.0002 & Germany & -- \\
\hline
Fix (Fiber) & 0.00002 & Germany & -- \\
\hline
Fix & 0.1 & Finland & 2020 \\
\hline
Fix & 0.03 & Europe & 2020 \\
\hline
Fix & 0.05 & Finland & 2022 \\
\hline
Fix (Fiber) & 0.04 & USA & - \\
\hline
\end{tabular}
\end{table}

Table~\ref{tab:fixenergyconsumption} presents data on energy consumption for fixed (wired) network communications across different locations and years, measured in kilowatt-hours per gigabyte (kWh/GB).

For fixed networks in Germany, \textcite{Prysmian2021} report an energy consumption of approximately 0.0002~kWh/GB for ADSL, compared to 0.00002~kWh/GB for fiber optic. Data for the USA from \textcite{Thundersaid2024} show that fixed fiber networks consume 0.04~kWh/GB. In Finland, \textcite{Pihkola2018} indicate a higher consumption of 0.1~kWh/GB for fixed networks. Data for Europe, provided by \textcite{EuropeanCommission2020}, are showing that fixed networks consume 0.03~kWh/GB. Additionally, \textcite{Traficom2022} reports a consumption of 0.05~kWh/GB for fixed networks in Finland. 

While the energy cost of data transmission and carbon emission factors are documented, data are lacking for traffic generated during video calls. The following section details the experimental protocol established to collect these measurements and the subsequent calculations used to convert data volumes into a comprehensive carbon footprint.

\section{Methodology}

\subsection{Experimental Setup}

The hardware used in the experiment consisted of an HP Elitebook laptop with an Intel Core i7 processor~\cite{Hp} using Windows 10 operating system~\cite{Windows10}. For videoconferencing meetings, the software used was Microsoft Teams~\cite{Teams}. No other windows were open during the majority of the meetings, which means that data consumption is exclusively linked to the meeting itself. The meetings were conducted using a 4G connection via a cell phone. To ensure that data consumption was exclusively linked to the video conference, all other applications on the cell phone remained closed throughout the experiment.

\subsection{Data Collection}

At the start of each meeting, mobile data consumption for the month was recorded using the Red by SFR French operator application~\cite{Sfr}. The following data were then noted: the start time, the number of participants, the number of cameras switched on, as well as whether a participant was sharing his or her screen, and the content shared. Each time one of these variables changed during the meeting, the time of the change and the variable concerned were recorded.

At the end of each meeting, mobile data consumption was again recorded, enabling the difference from the initial measurement to be calculated. This gave the specific data consumption for each meeting. 

A total of 9 meetings were held, during between 30 to 60 minutes. Meetings were classified as “camera on” when the cameras were on for almost the entire duration, and as “camera off” when the cameras were off for the majority of the meeting.

The dataset created includes details on call participants, data consumption, call duration, and various screen sharing activities. The key metrics analyzed are:

\begin{itemize}
    \item Data consumption per minute with the camera on;
    \item Data consumption per minute with the camera off.
\end{itemize}

\subsection{Carbon Footprint Calculation Model}

To estimate the final carbon footprint of online meetings, we used a mathematical model to convert data consumption into GHG emissions. The carbon footprint per minute of meeting, denoted as $\mathit{CF}$, is calculated as follows:

\begin{equation}
\label{eq:model}
\mathit{CF} = \mathit{DT} \times \frac{1}{1024} \times \mathit{EI} \times \mathit{CI}
\end{equation}
where $\mathit{CF}$ is the carbon footprint ($\mathrm{gCO_2e/min}$), $\mathit{DT}$ represents the average data transfer rate ($\mathrm{MB/min}$), $\mathit{EI}$ is the energy intensity of the network ($\mathrm{kWh/GB}$), and $\mathit{CI}$ = 79.1, the carbon intensity of the French electricity mix provided by ADEME (2013) ($\mathrm{gCO_2e/kWh}$)~\cite{Baseempreinte}. The ratio $\frac{1}{1024}$ represents the data conversion rate, with 1024 being the number of MB contained in a GB.

Since energy intensity values ($\mathit{EI}$) vary significantly across existing studies, we adopted an interval-based approach. We calculated a range of impacts using the minimum and maximum values identified in the state of the art.

Although data have been found for electricity consumption per amount of data transferred on fixed networks ranging from 0.00002 to 0.1~kWh/GB, we're going to retain ARCEP's assumption that the amount of electricity consumed varies little according to the amount of data transferred \cite{Arcep2019} for the continuation of the study, while bearing in mind that it still has a small impact.

Therefore, the subsequent calculations and results will strictly apply the energy intensity factors ($EI$) associated with mobile data transfer.

\section{Results}

The comparative analysis of data consumption reveals a significant disparity between the two modalities studied. As shown in Table~\ref{tab:data_cons}, the quantity of data transferred doubles when the video is enabled.

\captionsetup{font=footnotesize,justification=centering,labelsep=period}

\begin{table}[htbp]
\caption{Average data consumption per minute.}
\label{tab:data_cons}
\centering
\begin{tabular}{|l|l|}
\hline
\textbf{Camera status} & \textbf{Data consumption (MB/min)} \\ 
\hline
On             & 18.07                              \\
\hline
Off            & 8.76                               \\ 
\hline
\end{tabular}
\end{table}

The results indicate that keeping the camera on leads to a 106\% increase (approximately 2x) in data consumption compared to audio-only meetings.

Even though it has been stated that the study is based on \textcite{Arcep2019} assumption that the amount of data transferred does not vary the amount of electricity consumed on the fixed network, it can still be noted that electricity consumption in kWh per GB is 2.6 to 6,000 times higher on the mobile network than the data found for the fixed network.

Using the French carbon emission factor provided by \textcite{Baseempreinte}, electricity consumption contributes to the emissions of 9.49 to 20.57~g of CO$_2$e per GB on mobile networks.

Assuming 1 GB = 1024~MB, this leads to emission factors values from 0.0093 to 0.020~g of CO$_2$e per MB.

Applying the mathematical model defined in (\ref{eq:model}), the carbon footprint ($\mathit{CF}$) of a meeting with the camera on ranges from 0.17 to 0.36 g of CO$_2$e per minute, compared to 0.08 to 0.18 g of CO$_2$e when disabling camera.

\section{Discussion}
The previous analysis indicates a statistically significant difference in data consumption between having the camera on and off during video calls. However, this analysis is preliminary and does not account for screen sharing activities or instances where the camera was switched on or off during the call.

Based on the mathematical model defined in (\ref{eq:model}), the carbon emissions of a one-hour video call with a camera to one without can be compared with several use cases.

\begin{table}[htbp]
\caption{CO2 emissions of different activities.}
\label{tab:co2emissions}
\centering
\begin{tabular}{|l|c|c|c|}
\hline
\textbf{Activity} & \textbf{1 minute} & \textbf{1 hour} & \textbf{1 unit} \\
\hline
 & \multicolumn{3}{c|}{in g CO$_2$e} \\
\hline
Call camera-on & 0.17 -- 0.36 & 10.2 -- 21.6 & -- \\
\hline
Call camera-off & 0.08 -- 0.18 & 4.8 -- 10.8 & -- \\
\hline
Watching Netflix & 0.6 & 36 & -- \\
\hline
GPT-3.5 request & -- & -- & 0.6 \\
\hline
Hot shower & 18 & 1 080 & 162 \\
\hline
\end{tabular}
\end{table}

Table~\ref{tab:co2emissions} provides a detailed comparison of CO$_2$e emissions associated with various activities, measured in grams. For video calls, having the camera on results in emissions ranging from 10.2 to 21.6~g of CO$_2$e per hour. Turning the camera off reduces these emissions to 4.8 to 10.8~g of CO$_2$e per hour. Watching Netflix is shown to emit 0.6~g of CO$_2$e per minute, accumulating to 36~g of CO$_2$e per hour. A single request using GPT-3.5 generates 0.6~g of CO$_2$e. Additionally, taking a hot shower emits 18~g, 1,080~g, and 162~g of CO$_2$e per minute, per hour, and per unit, respectively~\cite{Carbonfootprintmodelingtool}.
Based on this data, a one-hour video call with a camera on is equivalent to about a one-minute hot shower or a 30-minute Netflix session.

Our findings suggest that turning the camera off during a one-hour meeting saves 5.4 to 10.8~g of CO$_2$e, which represents a saving of 50\% of the carbon footprint. 

It is then advisable to systematically use fixed networks such as Ethernet or Wi-Fi whenever possible. The quantity of data transferred does not significantly impact electricity consumption on these networks. For calls with a small number of participants, turning the camera on can foster social connection and happiness~\cite{Sherman2013}, which in turn can enhance motivation and productivity. However, during large meetings or webinars where active participation is not required, it is recommended to turn the camera off. Even though the impact is not significant, the quantity of data transferred has a small effect on electricity consumption on these networks.

\section{Conclusion}

The preliminary analysis indicates that turning the camera off during video calls can halve data consumption and significantly reduce carbon footprint, particularly on mobile networks. While further analysis is needed to account for screen sharing and dynamic camera usage, the current recommendations emphasize the use of fixed networks and judicious camera usage to optimize data consumption and electricity usage, therefore carbon footprint. When using mobile networks such as 3G, 4G, or 5G, it is best to turn the camera off unless it is essential to see each other. Even in environments where visual interaction is important, it is crucial to consider the impact on data consumption and electricity usage.

To provide a more comprehensive analysis, it is recommended to incorporate screen sharing data by analyzing the impact of different types of screen sharing, such as PowerPoint, Excel, Internet, and Code, on data consumption. This involves determining if screen sharing activities significantly affect data consumption when the camera is on versus off. Additionally, accounting for instances where the camera is switched on or off during the call will allow for a more detailed analysis to understand the dynamic impact of camera usage on data consumption. Furthermore, expanding the dataset by collecting additional data will strengthen the statistical significance of the results.

\section{Acknowledgments}

We would like to thank Boris RUF for his careful proof-reading and invaluable advice.

\printbibliography
\end{document}